\documentclass[preprint]{aastex631}
\usepackage{graphicx}
\usepackage{url}                                                              
\usepackage[]{natbib}     
\usepackage{epsfig}
\usepackage{color}
\usepackage{xspace}

\definecolor{DarkGreen}{rgb}{0.0,0.4,0.0}  
\newcommand{\B}[1]{{\color{blue}{#1}}}
\newcommand{\R}[1]{{\color{red}{#1}}}
\newcommand{\G}[1]{{\color{DarkGreen}{#1}}}
\newcommand{\M}[1]{{\color{magenta}{#1}}}

\usepackage{tikz}
\usetikzlibrary{tikzmark}

\usepackage[normalem]{ulem}
\usepackage{soul}
\usepackage{cancel}

\submitjournal{ApJ}
\shorttitle{Effects of supra-arcade downflows interacting with the post-flare arcade}
\shortauthors{Awasthi, Liu \& Gou}
\received{.......}
\revised{.......}
\accepted{......}

\begin{document}
	\title{Effects of supra-arcade downflows interacting with the post-flare arcade}
	
	\correspondingauthor{Arun Kumar Awasthi, Rui Liu}
	\email{arun.awasthi.87@gmail.com, rliu@ustc.edu.cn}
	
	\author[0000-0001-5313-1125]{Arun Kumar Awasthi}
	\affiliation{Space Research Centre, Polish Academy of Sciences, Bartycka 18A, 00-716 Warsaw, Poland}
	\affiliation{CAS Key Laboratory of Geospace Environment, Department of Geophysics and Planetary Sciences, University of Science and Technology of China, Hefei 230026, China}
	
	\author[0000-0003-4618-4979]{Rui Liu}
	\affiliation{CAS Key Laboratory of Geospace Environment, Department of Geophysics and Planetary Sciences, University of Science and Technology of China, Hefei 230026, China}
	\affiliation{CAS Center for Excellence in Comparative Planetology, Hefei 230026, China}
	\affiliation{Mengcheng National Geophysical Observatory, School of Earth and Space Sciences, University of Science and Technology of China, Hefei 230026, China}
	
	\author[0000-0003-0510-3175]{Tingyu Gou}
	\affiliation{CAS Key Laboratory of Geospace Environment, Department of Geophysics and Planetary Sciences, University of Science and Technology of China, Hefei 230026, China}
	\affiliation{CAS Center for Excellence in Comparative Planetology, Hefei 230026, China}

	\begin{abstract}
		Supra-arcade downflows (SADs) are dark voids descending through plasma above the post-flare arcade. Although they are generally viewed as byproducts of flare reconnections in the corona, the nature of SADs is under debate. Here we investigated six distinct episodes of SADs observed in the post-maximum phase of an M-class flare of April 11, 2013. Differential emission measure analysis revealed that SAD cases occurring close to the flare maximum contain an enhanced hot plasma component at 5--7~MK whereas those occurring later exhibited a depression in hot plasma at 7--12~MK compared to the ambient supra-arcade plasma. On-disk location of the flare enabled us to examine in detail the interaction of SADs with the post-flare arcade, whose effects include 1) transverse oscillations of period $\sim\,$160~s in the supra-arcade rays in the wake of voids, 2) footpoint brightening in 1700{~\AA} whose peak is delayed by 22-46~s with respect to the SAD's arrival at the top of the arcade, and 3) EUV intensity perturbations expanding and propagating with a speed $\sim\,$400 km~s$^{-1}$. On the other hand, due to line-of-sight confusion in the optically thin corona, the ribbon enhancement following the interaction produces an illusion of plasma rebound at the top of the arcade, where the interaction fails to yield significant plasma heating. These effects indicate that the interaction mainly generates MHD waves propagating toward the surface, which may further produce quasi-periodic brightening at flare ribbons, therefore contributing to quasi-periodic flare gradual phase emission in EUV.
	\end{abstract}
	
	
	\section{Introduction}
	Magnetic reconnection is the fundamental mechanism that energizes the solar eruptive processes. Since the direct observational characterization of magnetic reconnection is difficult \citep{Xue2016}, indirect proxies namely current sheets \citep{Sui2003, Liu2010, Gou2019}, loop shrinkage \citep{Forbes1996, Wang1999, Savage2010}, reconnection inflows and outflows \citep{Takasao2012, Su2013}, etc. are investigated for this purpose. Alongside the reconnection generated plasmoids (or blobs) which appear \emph{bright} in the extreme ultra-violet (EUV) \citep{Liu2013, Awasthi2018, Gou2019} and X-ray \citep{Milligan2010} wavelengths, careful examination of the supra-arcade region during the post-maximum phase of flares also reveal tadpole-shaped \emph{dark} voids descending towards the post-flare arcade, termed as supra-arcade downflows \citep[SADs;][]{McKenzie1999}. Unlike the bright plasmoids which are mainly observed during the flare maximum, multiple episodes of SADs are observed during the impulsive and main phases \citep{Asai2004} as well as throughout the prolonged decay phase of the flare \citep{Xue2020}. Therefore, probing the relatively lesser-known thermodynamical and magnetic nature of the SADs can offer new insights into the plasma heating process, particularly during the gradual phase of the flare.
	
	SADs were first identified as blob-shaped depressions in the X-ray intensity in the \emph{YOHKOH} observations \citep{McKenzie1999} that descend sunward with a speed generally ranging between 45 to 500 km--s$^{-1}$\citep{McKenzie2000, Asai2004}. The propagation speed of SADs being lower than the ambient Alfv\'en speed \citep[$\sim$1000 km--s$^{-1}$;][]{Verwichte2005}, as well as the free-fall speed, makes their magnetic characteristics elusive. To explain this behaviour, a drag force is proposed by \citep{Scott2013}. Still, as a few SAD cases were found to be spatio-temporally associated with the retracting flux-tubes, it has been argued that the SADs correspond to the cross-section of the flux-tubes \citep{McKenzie2000, McKenzie2009, Savage2010}. However, observations reveal that voids are systematically wider than the cross-section of the retracting loops. \citet{Savage2012} reconciled this discrepancy by re-interpreting the SADs as the `wakes' of the retracting loops. Alternatively, \citet{Liu2013} invoked a strong magnetic twist to account for both the size discrepancy between SADs and retracting loops and the depressed density in the SADs. Although the observed causal relationship of the retracting loops with the SADs in a few cases led the investigators to include the retracting flux-tubes \citep{Scott2013} to model the characteristics of the downflowing voids, it is not always the case. \citet{Liu2013} reported a continuous association between the downflowing voids and the retracting flare loops and inferred the former to be initiating the latter. Further, since the voids are estimated to contain a lower density ($\sim$ $10^9~cm^{-3}$) plasma compared to ``haze" in the supra-arcade region \citep{McKenzie1999, Reeves2017, Chen2017}, it is unclear how the rarefied region is developed behind the shrinking loops, which move with a speed smaller than the Alfv\'en speed \citep{Liu2013, Cassak2013}. Further, a few simulation studies, e.g., \citet{Maglione2011}, and \citet{Cecere2012}, considered plasma temperature in the SAD interior to be higher than that of the surroundings. Therefore, it is crucial to investigate the plasma as well as magnetic-field properties of the SADs as derived from the observations as well as from the models.
	
	Despite the complex nature of the observational characteristics exhibited by SADs, their co-temporal presence with the impulsive non-thermal bursts is indicative of their association with the magnetic reconnection occurring higher up in the corona \citep{Asai2004, Khan2007}. \citet{Liu2013} argued a common origin of the bright plasmoids in a vertical current sheet (VCS) generated in the wake of an erupting flux rope and a dark void flowing in the flare arcade. He conjectured the dark voids to be a mini-flux rope located in the stretched flux tubes of the VCS. On the other hand, the model of \citet{Cassak2013} revealed these low-density regions to be formed by continuous reconnection outflow jets that carve into the denser supra-arcade region. Additionally, \citet{Liu2021} noted that supra-arcade rays often exist prior to and/or are standalone, to the downflow events. As the prominence plumes \citep{Berger2010, Awasthi2019} and SADs exhibit morphological similarity, for instance, the splitting of a downflowing void head \citep{Innes2014}, Rayleigh-Taylor (R--T) instability in the downstream region of the reconnecting current sheet \citep{Guo2014} is suggested to be a promising mechanism of SADs. In this context, a mixture of instabilities, namely Kelvin-Helmholtz (K--H) \& the tearing mode \citep{Cecere2015}, or the Rayleigh–Taylor \& the Richtmyer-Meshkov \citep{Shen2022} have also been argued for the generation of SADs.
	
	The observational fact that SADs pass through, push, or oftentimes even split the structures in the supra-arcade region is indicative of their magnetic nature. \citet{Verwichte2005} interpreted the transverse oscillations generated in the trailing region of the SADs to be the propagating fast magneto-acoustic kink waves. Thus, although the interaction of the SADs with the structures in the supra-arcade region and later with the loop arcade has the potential of unraveling the physical nature of the downflows, it is not well-studied because most SADs are preferentially observed in the limb flares \citep{Savage2011}. Recently, \citet{Samanta2021} investigated a series of SADs, that occurred during the post-maximum phase of an on-disk M-class flare on April 11, 2013. Focusing their investigation on the collision of descending tadpoles with the post-flare loops, they found quasi-periodic heating of the loop plasma. At least six distinctively clear SAD events occurred throughout $\sim$~80 minutes during the decay phase of the flare, from which a subset of cases has been investigated in the above paper. To further the understanding of the nature of the interaction of SADs with the features in the supra-arcade region, post-flare loops, and foot-point regions, we carry out an exhaustive analysis of the dynamical and thermal properties of the SADs and their evolution in the present work. Novel results include the presence of hot plasma enclosed in the downflows observed close to the flare-maximum (section~\ref{sec:thermal_nature}) and the characterization of the enhancements and disturbances generated due to the interaction of SADs with the post-flare loop-arcade (section~\ref{sec:sad_osclln}). Discussion on the results and conclusions are offered in the section ~\ref{sec:disc_concl}.
	
	\begin{figure}
		\centering
		\includegraphics[width=0.9\textwidth]{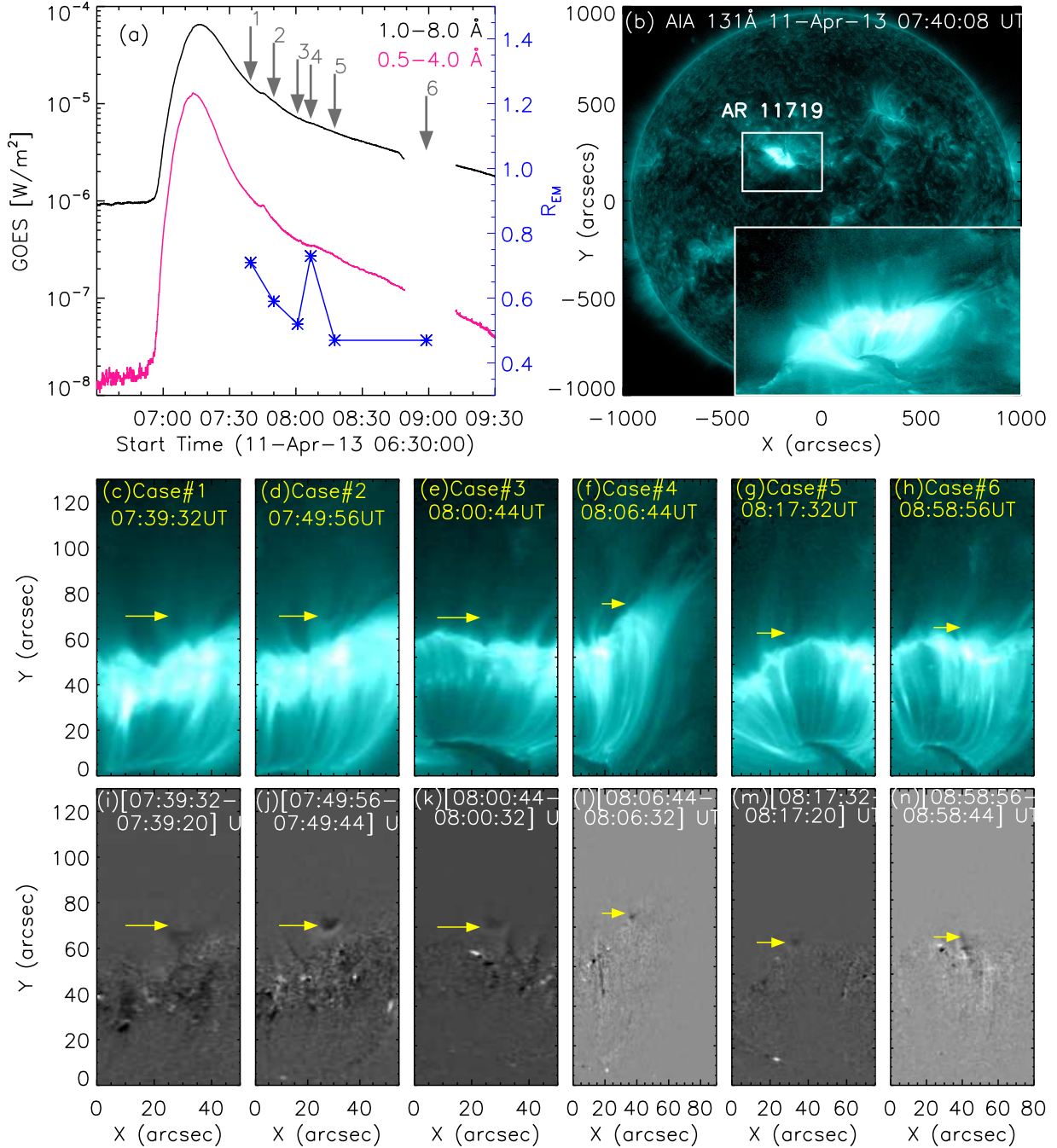}
		\caption{EUV overview of the SAD cases observed in the gradual phase of an M6.5 flare of April 11, 2013. (a) Disk-integrated X-ray intensity profile in 1--8{~\AA} (black) and 0.5--4{~\AA} (pink), depicting the prolonged gradual phase of the flare. Depression in EM values within SADs in the 7--12 MK temperature range is plotted in blue for all the SAD cases (scaled as right Y-Axis; discussed in section~\ref{sec:thermal_nature}). (b) Full-disk view of the flaring region AR 11719 in the 131{~\AA}. An upright view of the flaring region, prepared by zooming in and rotating the field-of-view of the active region by 130$^\circ$ clockwise is shown in the inset in panel (b). In addition to the post-flare loops, the bright supra-arcade region exhibiting various structures such as rays, dark lanes, and haze can be identified. Six cases of dark voids seen in the AIA 131~{\AA} images (denoted by yellow arrows) are shown in the middle panels ((c)--(h)) while corresponding running-difference images are plotted in the bottom panels ((i)--(n)), respectively. Images are rotated clockwise with angles varying in the range of 120--135$^\circ$ to present the upright view of the descending voids. SAD cases \#1--6 are annotated in panel (a) and also in the middle row of the figure. }\label{fig:sad_overview}
	\end{figure}
	
	\section{Overview of observations and Instruments}
	We investigate the dynamical and thermal characteristics of SADs that are observed in the post-maximum phase of an M6.5 flare of April 11, 2013, in the active region 11719. Primarily, we have analyzed the extreme ultra-violet (EUV) and ultra-violet (UV) images recorded by the Atmospheric Imaging Assembly \citep[AIA;][]{Lemen2012} instrument onboard \textit{Solar Dynamics Observatory} \citep[SDO;][]{Pesnell2012}. AIA provides un-interrupted observations of full-disk Sun in seven EUV channels (94, 131, 171, 193, 211, 304, and 335{~\AA}) and two UV wavelengths (1600 and 1700{~\AA}) with a spatial and temporal cadence of 0.6" and 12 s (24 s for UV channels), respectively \citep{Lemen2012}. The flare maximum occurred at 07:16 UT according to the disk-integrated X-ray intensity profile obtained from the \emph{Geostationary Observational Environmental Satellite} \citep[GOES;][]{Pesnell2012} in 1--8 {~\AA} and 0.5--4 {~\AA} (Figure~\ref{fig:sad_overview}a). The prolonged gradual phase of the flare (Figure~\ref{fig:sad_overview}a)) with its source region located close to the disk center (N10W01,
	Figure~\ref{fig:sad_overview}b) allowed a unique opportunity to identify at least six distinctively clear cases of SADs observed over 80 minutes (Figure~\ref{fig:sad_overview}c--h). Downflow events will be termed according to their observed sequence, i.e. SAD case\#1--6 (annotated in the figure~\ref{fig:sad_overview}a and also in the middle row of the figure).
	
	In particular, EUV images in 131 $\mathrm{\AA}$ offer the clearest visibility of the bright ``haze" in the supra-arcade region, and dark voids against brighter background (Figure~\ref{fig:sad_overview}b). Therefore, the identification and investigation of the morphological evolution of the dark voids have been made primarily using the image sequence obtained in 131 $\mathrm{\AA}$. The on-disk location of the flaring region enables us to observe the post-flare loop arcade and various structures in the supra-arcade region namely bright rays, dark lanes, haze, etc. (Figure~\ref{fig:sad_overview}b). However, running-difference images have also been prepared to deduce the morphological and kinematical properties of the descending voids (bottom panels of figure~\ref{fig:sad_overview}). This enabled us in identifying six cases of descending voids within 80 minutes of the decay phase of the flare, as marked by yellow arrows in Figure ~\ref{fig:sad_overview}a. For most cases, SADs descend toward the post-flare loops that are observed from a side-on perspective; Case \#4 is particular in that it descends toward cusp-shaped post-flare loops (figure~\ref{fig:sad_overview}f \& figure~\ref{fig:sad_overview}l), i.e., these loops are observed from an oblique face-on perspective. Thus, the curved post-flare arcade allows for an investigation of the effects of SAD interactions with the cusp-shaped loop structures in the supra-arcade region and with the loop-arcade from different geometrical viewpoints.
	
	\begin{figure}
		\centering
		\includegraphics[width=0.73\textwidth]{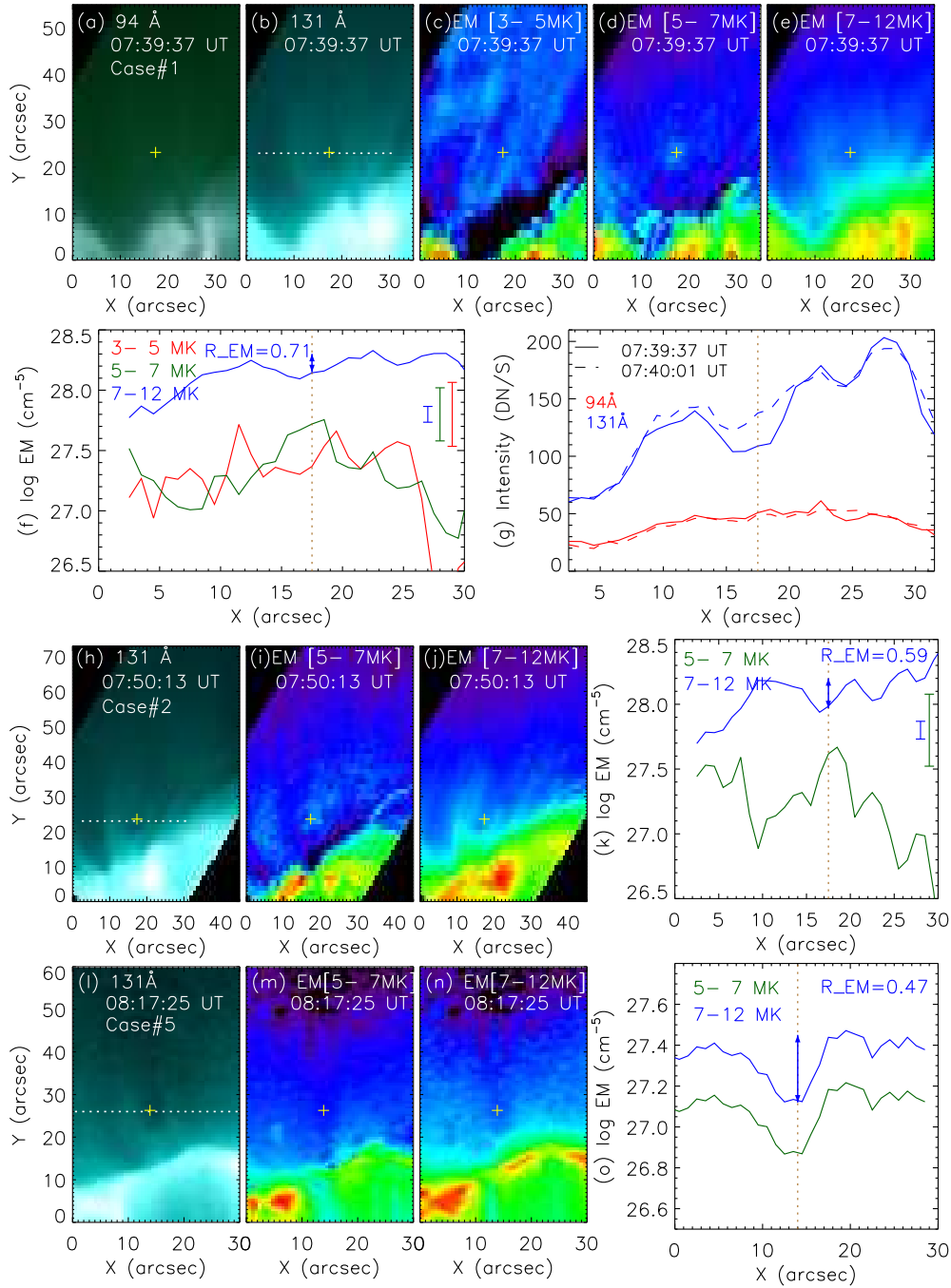}
		\caption{Thermal characteristics of three SADs as derived from the EUV observations. First two rows present thermal properties of SAD case\#1, observed at 07:39:37 UT, as observed in EUV wavelengths (94 and 131{~\AA}, shown in (a) \& (b)). Enhanced emission measure (EM) in 5--7 MK within the void is evident in the EM maps (3--5 MK, 5--7 MK, and 7--12 MK in panels (c)--(e)), as well as in the plot of EM values in various temperature ranges along a horizontal (yellow) line drawn in panel (b) slicing the blob. The maximum value of uncertainty in the respective EM values is plotted as error bars on the right. (g) 94{~\AA} (red) and 131{~\AA} (blue) intensity profiles along the horizontal line slicing the blob deduced from the EUV images acquired 07:39:37 UT (full) and 07:40:01 UT (dotted). `+' signs in EUV images and vertical lines in the plots mark the blob positions. Similarly, thermodynamical characteristics of SAD case\#2 (07:50:13 UT; panels (h)--(k)) also reveal the enhanced EM values in 5--7 MK temperature bin (EM map (i) \& plot (k)). (l)--(o) On the other hand, thermal properties of SAD case\#5 (at 08:17:25 UT), shown in the bottom panel (l--o), only exhibits depleted EM values at the location of void in all the investigated temperature bins. The level of depression in EM values (R\_EM) at the SAD location is shown as vertical double-headed arrow and annotated for SAD cases 1 (f), 2 (k), and 5 (o).}\label{fig:sad_thermal_properties}
	\end{figure}
	
	\section{Results} \label{sec:results}
	
	\subsection{Thermal properties of the supra-arcade downflows}\label{sec:thermal_nature}
	We deduce the thermodynamical properties of the plasma contained in the supra-arcade downflows in terms of its temperature and emission measure (EM). It is still debated whether the SADs contain plasma of lesser density (and higher temperature) compared to the ambiance or are devoid of plasma. The present investigation of multiple SAD cases spanned across $\sim$80 minutes since the first discernible episode (observed within a few minutes after the flare maximum), provides a unique opportunity to probe the thermodynamical nature of SADs that are produced with different flare reconnection rates --- most likely at different altitudes and therefore have cooled over different time periods at the time of their arrival at the post-flare arcade.
	
	To determine the thermal characteristics of the features observed in the supra-arcade region, we apply the modified sparse inversion technique of \citet{Su2018}, originally developed by \citet{Cheung2015}. This method analyzes the pixel intensity from the six EUV channels of AIA namely 94{~\AA}, 131{~\AA}, 171{~\AA}, 193{~\AA}, 211{~\AA}, and 335{~\AA} to determine the best-fit emission measure (EM) distribution over temperature. Besides, the implementation of Monte-Carlo simulation enables determining the uncertainty in the EM values derived from this technique. Figure~\ref{fig:sad_thermal_properties} provides the thermodynamical characteristics of three SAD cases, case\#1, 2 and 5, observed respectively at 07:39:37 UT, 07:50:13 UT, and 08:17:25 UT. From the investigation of case\#1, although a clearer perspective of the `haze' above the post-flare loops is seen in 131{~\AA} compared to that in the 94{~\AA} image, the post-flare loop region appears clearer in the latter wavelength. Further, EM maps have been prepared in three temperature bins (panel (c)--(e)), corresponding to low (3--5 MK), intermediate (5--7 MK), and high temperature (7--12 MK). Clear signatures of hot material within the blob may be noted in the EM map corresponding to 5--7 MK. In agreement, a plot of EM values (Figure~\ref{fig:sad_thermal_properties}f) along the horizontal line crossing the blob also shows an enhancement of EM only in the 5--7 MK range at the SAD's location. To account for the uncertainty involved in the identification of the SAD's position, EM value on each point along the horizontal line is estimated by taking an average of EM values in 3-pixels spanned vertically and centered on the respective point. Further, the statistical significance of the enhancement has been established by deriving the maximum uncertainty level in the EM values over the plotted range (vertical bars on the right-side of figure~\ref{fig:sad_thermal_properties}f). We have determined the intensity profile in 94 and 131{~\AA} wavelengths over the aforesaid horizontal line at two different times -- one during the passage of SAD (at 07:39:37 UT), and the other after its passage (at 07:40:01 UT). From the profile, we find elevated levels of emission in 131{~\AA} wavelength within the SAD, which may have significantly contributed to the estimated higher temperature of the SAD interior. Similar to thermal properties of aforementioned SAD, SAD case\#2, observed at 07:50:13 UT (Figure~\ref{fig:sad_thermal_properties}h--k), exhibited elevated EM values in the 5--7 MK temperature bin at the SAD location (Figure~\ref{fig:sad_thermal_properties}i \& k). On the contrary, thermal diagnostics of another SAD case\#5, observed at $\sim$ 08:17 UT (bottom panel of figure~\ref{fig:sad_thermal_properties}), did neither exhibit any discernible signature of enclosed hot material in the EM maps nor revealed any enhancement in the EM values within the SAD. However, we found a trend of increasing level of depression in the EM values ($R_{EM}$) at the SAD location in the 7--12 MK temperature range relative to that of the nearest vertical ray for the cases that occur early in the decay phase compared to those observed later. $R_{EM}$, defined as the ratio of EM values at the SAD location with that of the  nearest vertical bright supra-arcade ray, is estimated to be 0.71, 0.59, 0.52, 0.73, 0.47, and 0.47 for the SAD cases 1 to 6, respectively. Figure~\ref{fig:sad_thermal_properties} presents the analysis of SAD cases 1, 2 and 5, whereas the thermal characteristics of the rest of the cases in presented in the figure~\ref{fig:sad_thermal_properties_appendix} (see appendix~\ref{sec:appendix_1}). This suggests that the high-temperature plasma content within SADs is increasingly reduced for the cases that occur sequentially later in the decay phase (\textit{c.f.} figure~\ref{fig:sad_overview}a) except for that corresponding to SAD case \#4. The side-on view of the SAD\#4 and the presence of ambient bright cusp-shaped loop system allowed to discern the void only as late as very close to the bright post-flare looptop. Further, the EM profile within the SAD\#4 in the high-temperature bin (figure~\ref{fig:sad_thermal_properties_appendix}o) shows an unusual behaviour compared to a typical behaviour of EM depression within the SAD for the rest of the investigated cases. This may possibly explain why R\_EM for the SAD\#4 is an outlier. 

	\subsection{Investigation of emission and perturbation generated due to supra-arcade downflows}\label{sec:sad_osclln}
	To probe the magnetic field characteristics of descending voids, we investigate the disturbances they produce in the ambient supra-arcade structures (i.e. rays, cusps, etc.) and the post-flare loop arcade.
	
	\subsubsection{SAD generated transverse oscillations in the supra-arcade rays}\label{sec:sad_transverse_osclln}
\begin{figure}
		\centering
		\includegraphics[width=0.9\textwidth]{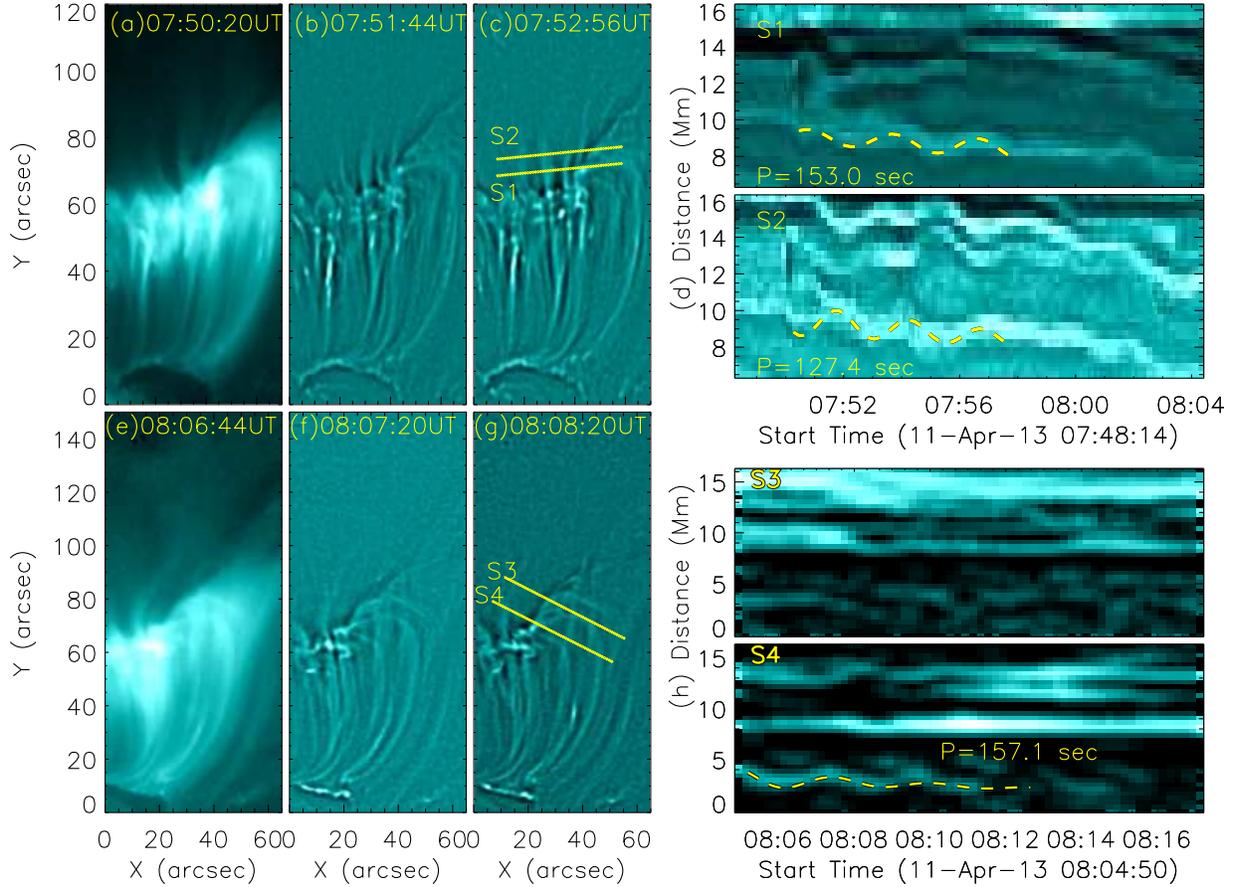}
		\caption{Transverse oscillations exhibited by supra-arcade rays in response to the passage of SAD cases. Top panel (a)--(d) shows the sequence of images in 131{~\AA} showing the passage of SAD case\#2 through the bright rays seen very clearly in the unsharp-masked images (b--c), and time-distance images (d) corresponding to virtual slits S1 and S2, placed across the rays as marked in panel (c). Identified oscillatory patterns are shown in yellow. The time-slice images show the transverse waves with periods ranging between 120-150 s. Similarly, the bottom panel displays the transverse waves produced by SAD case\#4 as seen in the 131{~\AA} images (e--g), and time-distance images (h) plotted along the slits S3 and S4 placed across the supra-arcade loops, shown in the panel (g). The period of the waves seen in the time-distance images is 157 s. An animation containing the sequence of images in AIA 131{~\AA}, unsharp-masked, and running-difference images, is made available online to show the intensity waves and perturbations generated due to the interaction of SADs with the supra-arcade rays.}\label{fig:SAD_ray_transverse_osclln}
	\end{figure}

	We have processed the 131{~\AA} images with the unsharp-mask technique to highlight the fine structures during the passage of voids. From the original images, we subtract a background image (the mask) that is a smoothed (7 $\times$ 7 pixels) version of the original image. During downward propagation, SADs have been observed to push aside the bright vertical supra-arcade rays in all of the investigated cases. However, we selected to show the detailed analysis of case\#2 and case\#4 in the top and bottom panels of figure~\ref{fig:SAD_ray_transverse_osclln}, respectively, since they provide different geometrical perspectives.
	
	Following the passage of the voids, the perturbed rays in the supra-arcade region exhibit an oscillatory behavior (figure~\ref{fig:SAD_ray_transverse_osclln}a--c \& e--g). We prepare time-distance maps corresponding to the virtual slits placed on the sequence of unsharp-masked 131{~\AA} images to quantify the oscillations. In the top panel corresponding to case\#2, slits `S1' and `S2' are placed (figure~\ref{fig:SAD_ray_transverse_osclln}c) across two discernible rays along which the void has descended. Time-distance maps corresponding to the slits exhibit the transverse nature of the oscillation of these rays. We fit the oscillatory pattern with the damped-sinusoidal function \citep{Awasthi2019} which reveals the period of oscillations for case\#2 to be varying in the range of 120--160 s.
	
	Similarly, we investigate the oscillatory behavior of the selected post-flare loops due to the passage of void\#4 as presented in the bottom panel of the figure~\ref{fig:SAD_ray_transverse_osclln}. From the sequence of unsharp-mask images at 08:07:20 UT and 08:08:20 UT (figure~\ref{fig:SAD_ray_transverse_osclln}f--g and the associated animation covering the entire descent of the void), we found that both sides of the cusp-shaped loop structure exhibited transverse oscillations following the passage of the void. That the cusp-shaped loops face toward us enabled us to identify the transverse oscillations, which may correspond to transverse oscillations of supra-arcade rays due to different viewing angles. The time-distance maps corresponding to slits `S3' and `S4', which span across one of the distinctively clear cusp structures, show the oscillations with a period of 157.1 s.
	
	The detected periods of oscillation in the supra-arcade rays seem to be consistent with the period of standing kink waves \citep{Aschwanden2002}. Further, the change of period at different altitudes of the supra-arcade ray may be related to the magnetic field in different rays or at different altitudes. This may also be caused by the superposition of numerous rays along the line of sight, but it would be difficult to resolve the differences from the EUV images. It may also result from large uncertainties involved in delineating and fitting the oscillatory patterns since for the same ray (flux tube), the oscillation period should be the same.
	
	\subsubsection{EUV intensity waves propagating in post-flare loops due to SAD interaction} \label{subsec:intensity_wave}
\begin{figure}
		\centering
		\includegraphics[width=0.9\textwidth]{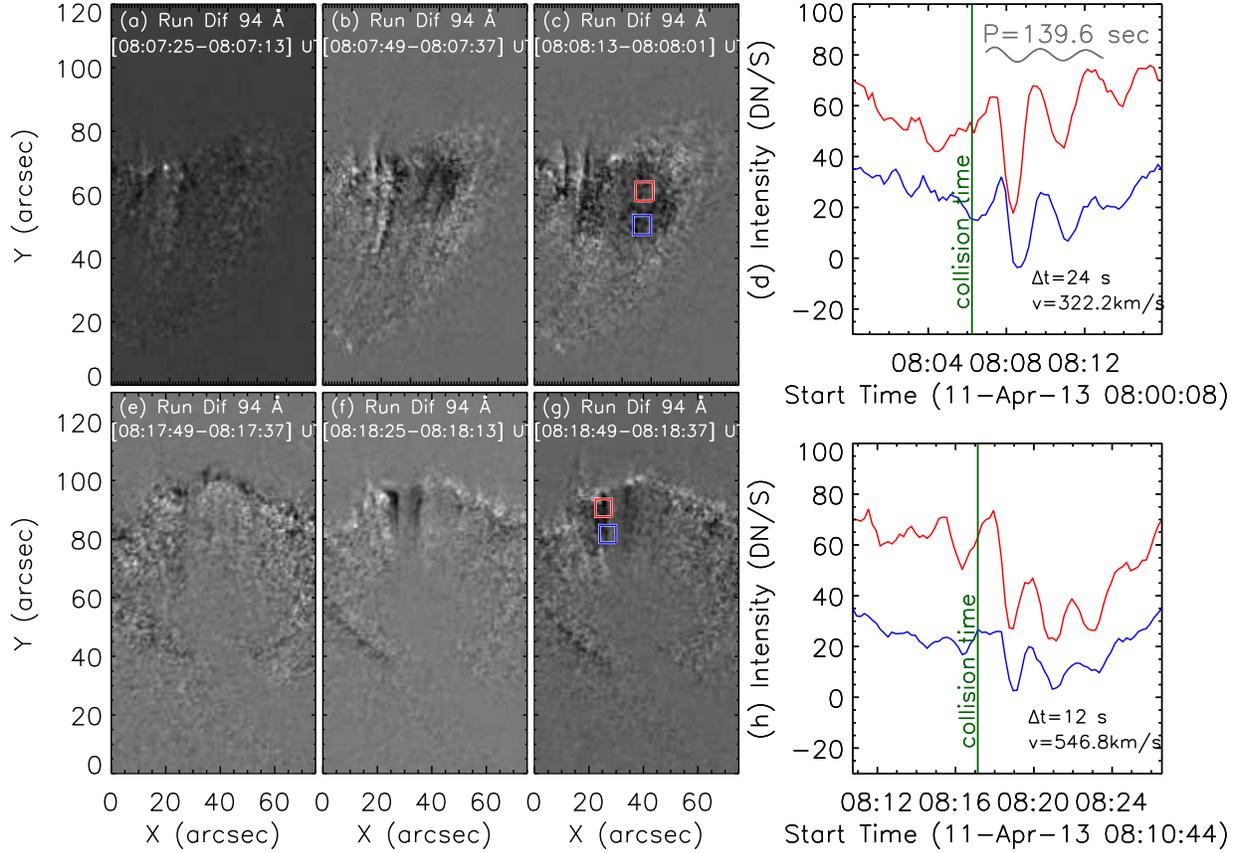}
		\caption{EUV intensity oscillations exhibited by the post-flare loop-arcade generated due to the collision of void case\#4 and case\#5, presented in the top and bottom panels, respectively. (a)--(c) Sequence of running difference images in the 94{~\AA} observations showing the generation and expansion of intensity perturbations. (d) Intensity profiles in 94{~\AA} as obtained by averaging the emission within two square regions 6" width, placed 10" apart (shown in panel (c)). Intensity oscillation corresponding to the bottom region (blue) is delayed by 24 sec (as determined from the lag analysis) from that obtained from the higher (red) region, thus revealing the propagation speed of oscillations along the loop to be 322.2 km--$s^{-1}$. The oscillation period is estimated to be 139.6 s, shown using a representative sinusoidal pattern plotted in grey. Similarly, the bottom panels show the sequence of running difference images (e--g) as well as intensity profiles in 94{~\AA} (h) corresponding to SAD case\#5. Here the propagation speed of intensity oscillations is estimated to be 546.8 km--$s^{-1}$. The vertical green line marks respective times of SAD's collision with the top of the post-flare loop-arcade.}\label{fig:sad_intensity_perturbation}
	\end{figure}

	The on-disk location of the flare's source region provided an opportunity to investigate the effect of SADs' interaction with the post-flare loops, an aspect not well known. A sequence of running-difference images in 94 and 131{~\AA} wavelengths have been prepared for this purpose. A detailed investigation made on the SAD cases\#4 and case\#5 is presented in figure~\ref{fig:sad_intensity_perturbation}.
	
	The passage of SAD case\#4 over the post-flare loop-arcade resulted in a downward propagating and expanding region of intensity depression in the running-difference images. These wave-like perturbations in the EUV intensity profiles are deduced by averaging the intensity over two square regions of 2" width, placed 10" apart (marked in figure~\ref{fig:sad_intensity_perturbation}c). Thus, the intensity profiles obtained from the two regions along the loops enabled us to deduce the time delay between the wave patterns (24 s as determined by performing lag analysis), and hence the propagation speed of intensity waves, which is estimated to be about 322 km/s. All the SAD cases under investigation produced such intensity waves in the post-flare arcade, as demonstrated for another SAD case in the lower panel of figure~\ref{fig:sad_intensity_perturbation}. In this case, the wave generated due to the collision of the SAD with the post-flare loops has propagated along the loop with a speed of 547 km--$s^{-1}$. The period of oscillations is estimated to be $\sim$140 s.
	
	\subsubsection{Investigation of loop-top and foot-point brightenings associated with the descending voids}\label{sec:sad_ribbon_reponse}
\begin{figure}
		\centering
		\includegraphics[width=0.9\textwidth]{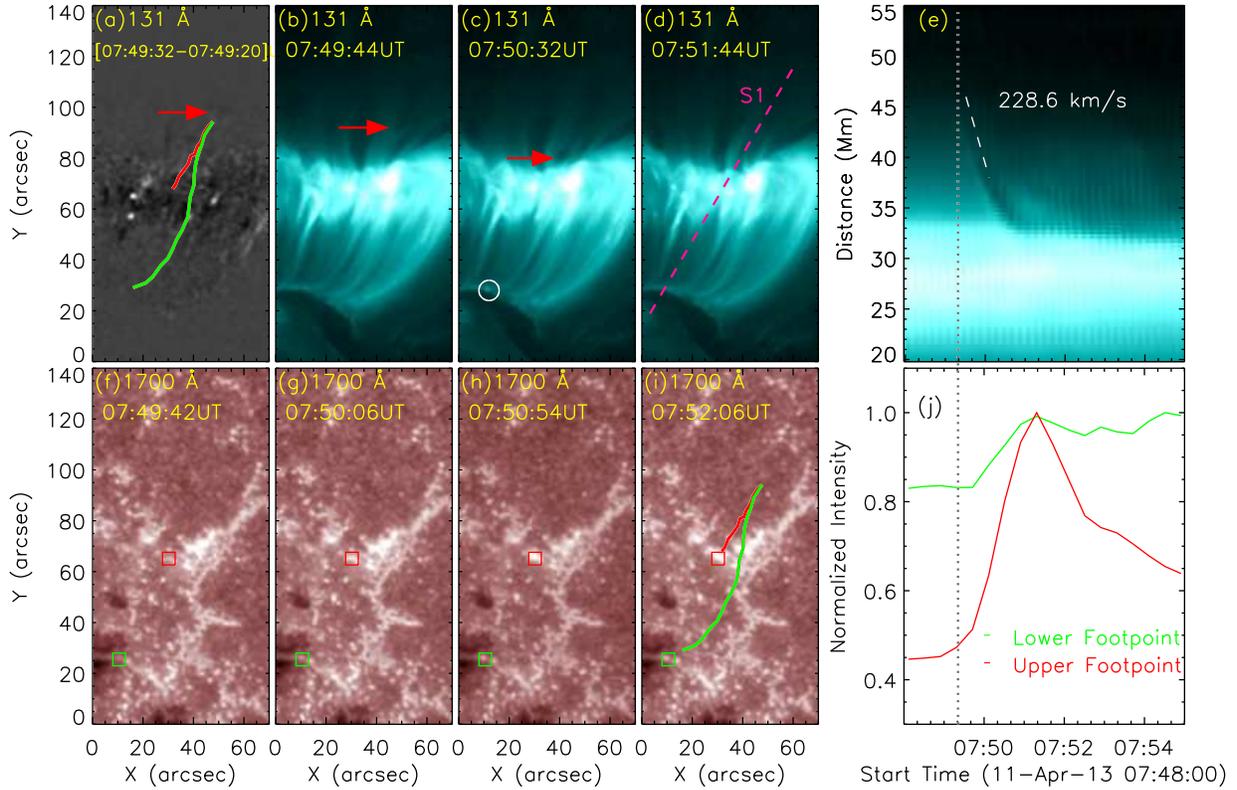}
		\caption{EUV and UV enhancements at the loop top and foot-points due to SAD's \textbf{(case\#2)} interaction with the supra-arcade rays. A sequence of images in 131{~\AA} (a--d) covering the SAD's propagation since its first discernible appearance (running-difference image; (a)) until it submerged below the post-flare loop-arcade. The respective closest-in-time 1700{~\AA} images are shown in panels (f)--(i). Envisaged loop structure (front (green) and back (red)) relevant to the SAD's interaction is drawn in panels (a) and (i). Red arrows in panels (a)--(c) mark the location of SAD whereas the associated lower (upper) foot-point brightenings in the 1700{~\AA} images are marked in green (red). Time-distance map in 131{~\AA} wavelength (panel (e)) corresponding to the virtual slit `S1' drawn in pink in panel (d). (j) Intensity profiles averaged over 5" wide square box from the sequence of images in 1700{~\AA} corresponding to lower and upper foot-points are plotted. The vertical dotted line (at 07:49:32 UT) in the time-distance map and intensity plot marks the time of SAD's first discernible appearance in the 131{~\AA} images. An animation is made available online which contains the sequence of images in 131, 1700 and 1600{~\AA} wavelengths, time-distance map in 131{~\AA}, and ribbons' intensity evolution in 1700{~\AA}.}\label{fig:sad_propagation_to_looptop_case2}
	\end{figure}

	Probing the spatio-temporal and thermodynamical characteristics of the emissions generated due to the interaction of SADs with the cusp-shaped loops in the supra-arcade region and their subsequent collision with the post-flare loops can provide crucial insights into the physical nature of SADs. In this regard, we investigate an SAD event (case\#2) which was observed at $\sim$07:49 UT (figure~\ref{fig:sad_propagation_to_looptop_case2}). To track the propagation of downflows since their earliest appearance in the supra-arcade region and identify the associated footpoint brightenings, sequences of images in 131 and 1700{~\AA} have been analyzed. Against the bright hazy background in the supra-arcade region, which is assumed to be associated with the cusp-shaped loops overlying the post-flare arcade from this side-on viewing angle, the investigated void could be identified in the 131{~\AA} images as early as at 07:49:32 UT (figure~\ref{fig:sad_propagation_to_looptop_case2}a--d). Intriguingly, merely 10 seconds after the interaction of the SAD with the loop structure, a kernel-shaped enhancement has been located in the lower ribbon in the 1700{~\AA} image acquired at 07:49:42 UT (figure~\ref{fig:sad_propagation_to_looptop_case2}f). It may be noted that although the time-cadence of 131 and 1700{~\AA}  images is 12 and 24 seconds, respectively, the time difference between the closest-in-time inter-wavelength images can be shorter, as in this case. From an animation containing the sequence of images in 131, 1700, and 1600{~\AA} wavelengths (supplementary file of figure~\ref{fig:sad_propagation_to_looptop_case2}), the spatial and temporal association of ribbon brightening with the SAD's dynamics can be further elucidated. For consistency, an additional criteria of selecting the closest-in-time 1700{~\AA} that are recorded later than the respective 131{~\AA} image has been implemented for investigating the causal relationship between the SAD's interaction with the loop and foot-point brightening. Kernel-shaped brightening is also observed along the upper ribbon.
	
	To derive a spatio-temporal relationship between the interaction of the SAD with the post-flare loop-arcade and the ribbon brightening, we prepare time-distance maps corresponding to a virtual slit `S1' (figure~\ref{fig:sad_propagation_to_looptop_case2}d\&i), placed in a slanted position to cover the kinematics of the descending void as well as the locations of ribbon enhancements. Both the upper and lower ribbons start to brighten as the SAD interacts with the supra-arcade rays as high as at $\sim$12 Mm above the post-flare loop arcade, and the enhancements peak as the SAD reaches the top of the post-flare arcade. From the shape and orientation of the post-flare loops, we envisaged a cusp-shaped loop structure relevant to the SAD's descent, with its foot-points rooted at the locations where brightenings are first spotted in the 1700{~\AA} images. At the foot-point locations, we further derive the time evolution of 1700{~\AA} emission (figure~\ref{fig:sad_propagation_to_looptop_case2}j) averaged over a 5" wide square box covering the ribbon enhancement (figure~\ref{fig:sad_propagation_to_looptop_case2}f--i). This also confirms that foot-point brightenings are associated with the SAD's descent.
	
\begin{figure}
		\centering
		\includegraphics[width=0.9\textwidth]{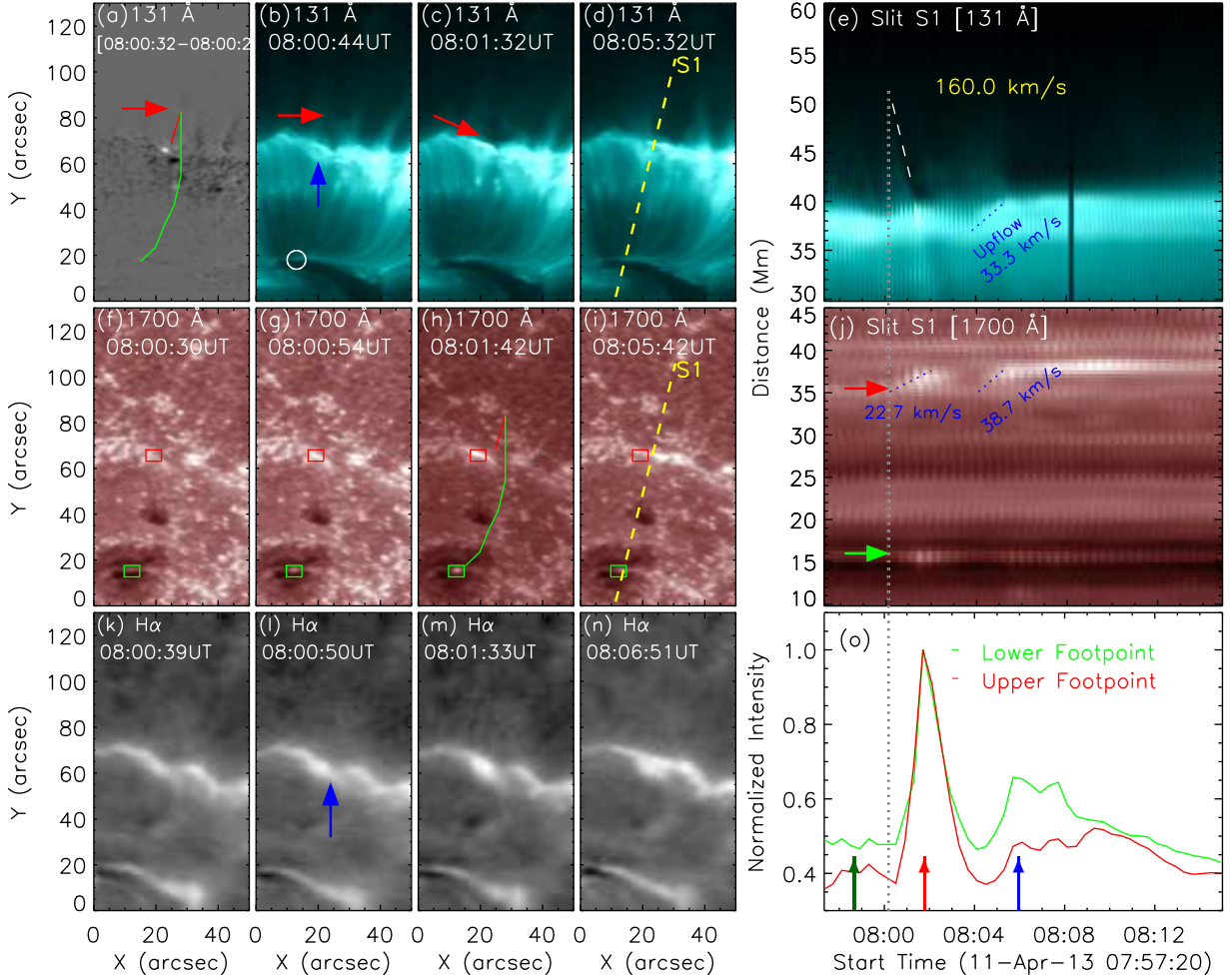}
		{\color{DarkGreen} \thicklines \put(-144,26){\vector(0,1){20}}}
		{\color{red} \thicklines \put(-121,26){\vector(0,1){20}}}
		{\color{blue} \thicklines \put(-89,26){\vector(0,1){20}}}
		\caption{EUV, UV, and H$\alpha$ emissions associated with the SAD's (case\#3) interaction with the post-flare loop-arcade. A sequence of images in 131 {~\AA} (a--d) covering the SAD's propagation from its first discernible appearance (running-difference image; (a)) until it submerges below the post-flare loop top. Respective closest-in-time 1700{~\AA} and H$\alpha$ images are shown in panels (f)--(i), and (k)--(n), respectively. Red arrows in the panels (a)--(c) indicate the location of SAD, and associated enhancement in the upper ribbon in the 131{~\AA} (b), and H$\alpha$ images (l) are marked in blue. Time-distance maps in 131{~\AA} (e) and 1700{~\AA} (j) wavelengths corresponding to slit `S1' drawn in pink in panels (d) and (i). SAD propagation and dynamical characteristics of brightenings at the ribbon and collision site are annotated. Upper (lower) foot-point location of the envisaged cusp-shaped loop structure relevant to the investigated SAD case (plotted in panels (a) and (h)) is marked with 5" wide square boxes in red (green), and the averaged intensity in 1700{~\AA} images corresponding to these regions is plotted in panel (o). The vertical dotted line at 08:00:10 UT displays the onset of emission enhancements at the ribbon location since the SAD's appearance in 131{~\AA}. Vertical arrows show the time instances for which the EM distribution has been derived in figure~\ref{fig:sad_case3_thermal_properties}. An animation containing the sequence of images in 131, 1700{~\AA} and in H$\alpha$ line-center wavelengths, time-distance map in 131{~\AA}, and ribbons' intensity evolution in 1700{~\AA} is made available online.}\label{fig:sad_propagation_to_looptop_case3}
	\end{figure}

	To further explore how the flare ribbons respond to the SAD's descent, we examine the nature of footpoint emission associated with another SAD case, observed at $\sim$08:00 UT (case\#3) by analyzing EUV observations in 131{~\AA} and 1700{~\AA} along with H$\alpha$ images obtained from Kanzelhoehe Solar Observatory (KSO). Distinctively clear kernel-shaped brightenings are seen at both the ribbons in the 1700{~\AA} images since the earliest discernible appearance of void in the 131{~\AA} images at 08:00:32 UT (figure~\ref{fig:sad_propagation_to_looptop_case3}a--b). The spatio-temporal relation of SAD's descent with the ribbon brightening is further elaborated with the help of the animation prepared from the sequence of images in the 131, and 1700{\AA}, as well as H$\alpha$ wavelengths. From the sequence of images in 1700{~\AA}, we further derive intensity profile (figure~\ref{fig:sad_propagation_to_looptop_case3}j) averaged over the foot-point region (5" wide square box, marked in green (red) for lower (upper) foot-point in figure~\ref{fig:sad_propagation_to_looptop_case3}f--i) of the envisaged loop structure affected by the SAD propagation (drawn in panels (a) and (i) of figure~\ref{fig:sad_propagation_to_looptop_case3}) and normalize to the respective peak value.
	
	Next, we prepare a time-distance map from the 131{~\AA} images over a virtual slit placed along the propagation path of the void (S1; drawn in figure~\ref{fig:sad_propagation_to_looptop_case3}d). New episode of enhanced emission at the foot-point location appears with a time-delay of $\sim$22 s following the SAD's appearance at the loop-apex in the 131{~\AA} image at 08:00:32 UT (figure~\ref{fig:sad_propagation_to_looptop_case3}a) which is located $\sim$ 15 Mm above the bright post-flare loop-top. Besides, the propagation speed of the SAD interaction to the footpoint can be more reliably derived from the peak time of the footpoint brightening relative to the time that the SADs arrive at the top of the post-flare arcade. In this regard, based on the investigation of SAD case\#3 presented in figure~\ref{fig:sad_propagation_to_looptop_case3} and associated animation, we obtain the arrival time of downflow at the post-flare looptop to be at 08:01:20 UT, while the intensity profile in 1700{\AA} gives the peak time of the footpoint brightening at 08:01:42 UT (figure~\ref{fig:sad_propagation_to_looptop_case3}h). Thus, from the estimated time delay of 22 s and the distance of 21.9 Mm from the loop-top to the footpoint, the propagation speed of perturbation is obtained to be $\sim$995 km s$^{-1}$. Similarly, we have also estimated the propagation speed of the perturbation for case \#2, presented in figure~\ref{fig:sad_propagation_to_looptop_case2}. Using the arrival time of the SAD at the top of the post-flare loop arcade to be at 07:50:56 UT, the peak emission at the ribbon location at 07:51:18 UT (time-delay = 22 s), and the distance of 25.5 Mm between the SAD's interaction site at the post-flare looptop and the ribbon brightening location, the speed of perturbation is derived to be $\sim$1159 km--s$^{-1}$. However, the estimation of time-delay between the SAD's arrival at the post-flare looptop and footpoint brightening may contain a maximum uncertainty of 24~s, due to the different time-cadence of images being recorded in 131~\AA (12 s) and 1700~\AA (24 s) images. Therefore, the uncertainty in the estimated speeds may be as large as 500 km--s$^{-1}$.
	
	From the careful analysis of the two cases of SADs' interaction with the loop system and its effect on the loop footpoints, it is evident that SADs start to interact with the flare loop system before they arrive at the top of the arcade. However, it is hard to determine exactly when the interaction starts because of the increasingly poor contrast between the SADs and the ambient corona with increasing height. Moreover, it is much more accurate to determine the peak than the onset of the ribbon brightening despite the fact that the latter depends on the instrument sensitivity and is interfered by the plage brightening.
	
	\subsubsection{Thermal analysis of SAD induced brightening at the collision-site and foot-point}
	Alongside enhanced kernel-shaped brightening at the lower foot-point region of the loop-system, the collision of SAD\#3 with the loop-top also resulted in localized plasma heating close to the collision site
	(figure~\ref{fig:sad_propagation_to_looptop_case3}b), which exhibited an apparent upward motion in the course of time (figure~\ref{fig:sad_propagation_to_looptop_case3}b--d). To investigate this association, we prepared a time-distance map (figure~\ref{fig:sad_propagation_to_looptop_case3}e) from the 131{~\AA} images corresponding to a virtual slit (`S1' shown in figure~\ref{fig:sad_propagation_to_looptop_case3}d). Here, the slanted dark lane is due to the void which descends with a speed of 230.8 km--s$^{-1}$ before getting decelerated near the loop-top. The void continued to be discernible even below the loop-top altitude before its disappearance. An upward motion of the bright material within a few minutes after the collision of the void is marked in blue in figure~\ref{fig:sad_propagation_to_looptop_case3}e. This is similar to the results obtained by \citet{Samanta2021} who interpreted the upward motion as the rebound of plasma that has been compressed due to SAD's collision.
	
	Since the upper ribbon is shadowed by the loop-top as evident from the 131{~\AA} and 1700{~\AA} images (figure~\ref{fig:sad_propagation_to_looptop_case3}b\&g), and the speed of ``upward" motion of the bright material observed in both the aforesaid wavelengths is estimated to be $\sim$35 km-$s^{-1}$ (figure~\ref{fig:sad_propagation_to_looptop_case3}e\&j), it is unclear whether the aforesaid hot material produced after the collision belongs to the loop-top region or the foot-points (ribbon). To resolve this ambiguity, the images acquired in 1700{~\AA} and H$\alpha$ wavelengths have been analyzed. The 1700{~\AA} and H$\alpha$ images (figure~\ref{fig:sad_propagation_to_looptop_case3}g\&l) acquired closest-in-time with the EUV 131{~\AA} image have revealed newly appeared enhancements co-spatial to that observed in the 131{~\AA} images (figure~\ref{fig:sad_propagation_to_looptop_case3}b; indicated by blue arrow). From the supplementary animation file provided with figure 6, and the time-distance map prepared from the sequence of images in 1700{~\AA} (figure~\ref{fig:sad_propagation_to_looptop_case3}e\&j) along the slit `S1' (same as that used in 131{~\AA} map), two episodes of ``upward moving" brightenings at the location of shadowed (upper) ribbon can be identified. The first instance, at 08:00:32 UT, is a response to the SAD’s interaction with the loop apex as evident by placing a vertical dotted line on the time-distance maps. Following the post-collision submergence of the SAD beneath the post-flare looptop, the second instance of brightening and ribbon expansion occurred at $\sim$08:04 UT, exhibiting similar spatial and temporal evolution as well as projected speed to that observed in 131{~\AA} time-distance map. Thus, it is very likely that ribbon brightening and expansion significantly contributes to the brightening and apparent upward motion observed in the 131{~\AA} images at the collision site. Contrary to the ``expansion" exhibited by the bright material in the upper-ribbon, the emission originated in the lower-ribbon remained relatively stationary, probably because the lower-ribbon runs across the sunspot, where the magnetic field is quite ``rigid".
	
\begin{figure}
		\centering
		\includegraphics[height=0.9\textwidth, angle=90]{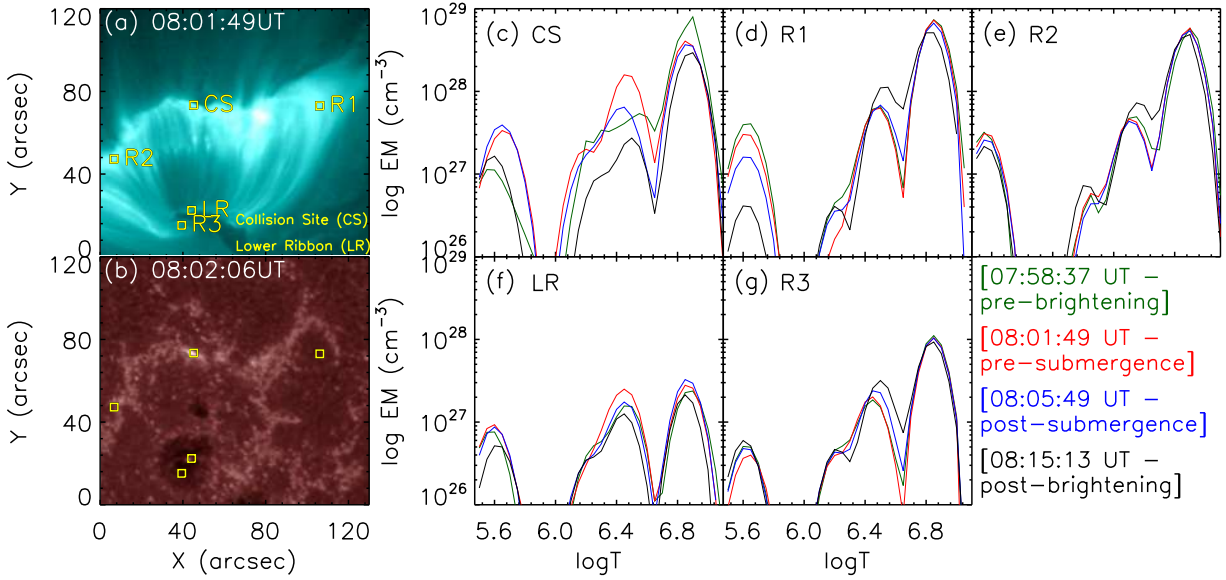}
		\caption{Thermal characteristics of the EUV brightenings generated due to SAD's interaction with the post-flare loop-arcade. Two episodes of E/UV enhancements in 131{~\AA} and 1700{~\AA} images have been observed pre- (08:01:49 UT; (a) \& (b)) and post-submergence of the SAD in the post-flare loop arcade. EM distributions at four time instances (two time instances representing pre- and post-submergence brightenings, and the other two representing before and after the brightening activities; annotated at the bottom-right and marked in the 1700{~\AA} intensity profile of the ribbon brightening plotted in figure~\ref{fig:sad_propagation_to_looptop_case3}o), averaged over a square box of 6 pixels width corresponding to enhanced emission region identified visually at collision-site (CS), two reference regions R1 \& R2 located at the top of the post flare loop arcade, at Lower Ribbon (LR), and at a reference point R3 which is at the footpoints of post-flare loops but off the LR, are plotted in panels (c) to (g), respectively. Note that the locations of CS and LR have changed in time due to the dynamic nature of the brightening, whereas the location for the reference points have been kept stationary. An animation containing the EM[T] distribution for the aforementioned locations during 07:56:49--08:20:13 UT is available online.}\label{fig:sad_case3_thermal_properties} 
	\end{figure}

	To further delineate whether the brightening at the collision-site corresponds to the looptop altitude or originates at the ribbon location, we investigate the thermodynamical characteristics of the EUV brightenings by analyzing the emission measure maps which are synthesized in the temperature range of 0.5--16 MK ($\log T$[K] = [5.5--7.2], $\delta\log T$=0.5) using the procedure described in section~\ref{sec:thermal_nature}. We deduce emission measure distribution (EM[T]) of the brightenings observed at the collision site (hereafter `CS'). Since the upper ribbon is shadowed by the post-flare loops, EM distribution corresponding to the enhancements at the lower (``exposed") ribbon (hereafter `LR') has also been deduced to distinguish the contribution of ribbon and looptop in the EUV enhancement observed at CS during SAD's collision. We determine the EM[T] distribution for three additional reference locations (refer to figure~\ref{fig:sad_case3_thermal_properties}) to determine the thermodynamical state of plasma -- at R1 and R2, which correspond to the top of post-flare arcade yet away from the region affected by the interaction of SAD (R1 at the cusp region, which is exactly the top of post-flare arcade without the plage emission along the line-of-sight; R2 is projected off the upper ribbon but coincident with brightening plages, providing a condition similar to that at CS) -- and at R3, which is located at the footpoints of post-flare loops but off the lower ribbon. EUV enhancements at CS (in 131{~\AA} images) and LR (in 1700{~\AA} images) locations have been observed as soon as the SAD interacts with the supra-arcade rays. 
	
	In particular, two episodes of EUV brightening at CS have been observed during 08:01 -- 08:11 UT during the descent and subsequent submergence of the SAD\#3 below the post-flare loop-top. To outline the effects of the SAD's interaction with the cusp of the post-flare loop-arcade and subsequent submergence on the thermodynamical state of the plasma, we deduce the EM[T] distributions at four time instances -- corresponding to two episodes of brightenings related with the SAD's pre-submergence (08:01:49 UT) and post-submergence (08:05:49 UT) below the post-flare loop-top, and two time instances before (07:58:37 UT), and after (08:15:13 UT) the brightening event (figure~\ref{fig:sad_case3_thermal_properties}, also see figure~\ref{fig:sad_propagation_to_looptop_case3}o). Due to the dynamical nature, the location of enhancements at CS and LR have been visually identified at each time-instance during 08:00:49 -- 08:11:13 UT for determining thermal characteristics whereas the locations of reference points, namely R1, R2, and R3, are kept fixed. Further, the CS and LR locations during the pre- and post-submergence time instances are kept the same as those identified at the onset (08:00:49 UT) and fading (08:11:13 UT) of the brightening, respectively. EM[T] distribution is derived by averaging the EM values over a square region of 6 pixels width from the emission measure maps in respective temperature bins. A comparison of EM[T] distribution of the enhancement at CS with that deduced before (07:58:37 UT) and after (08:15:13 UT) the brightening event (figure~\ref{fig:sad_case3_thermal_properties}c) reveal clear enhancements up to 5 MK ($\log T=6.7$), which is not evident at the reference points R1 and R2 (figure~\ref{fig:sad_case3_thermal_properties}d \& e). Besides, the EM[T] distribution in the hot bump ($\log T$[K] = [6.7--7.2]) did not show any significant change relative to that deduced at an earlier time as well as those at the reference points R1 and R2 (figure~\ref{fig:sad_case3_thermal_properties}d--e). Similarly, EM values at LR are found to be slightly higher than that corresponding to pre- and post-collision time (figure~\ref{fig:sad_case3_thermal_properties}f). Similar conclusions can be drawn from the EM[T] distributions corresponding to EUV enhancement observed after the submergence of SAD at 08:05:49 UT. 
	
	A detailed analysis of EM[T] distribution at CS revealed that the most prominent feature is not just the enhancement of EM[T] at the warm bump ($\log T=6.5$) but at the cold bump ($\log T=5.7$). In particular, EM[T] levels of CS at the cold bump consistently remained at an elevated level during the SAD's submergence event until 08:13 UT, after which the pre-collision values have been attained, whereas at the two reference regions the cold bump of EM[T] either keeps decreasing (R1) or remains stable (R2) during the same period (figure~\ref{fig:sad_case3_thermal_properties}d and animation). Thus, it can be argued that this enhancement at the cold bump is strong evidence for the contribution of enhanced ribbon emission in the transition region, due to Fe VIII ($\log T=5.6$) emission lines in the AIA 131~{\AA} passband \cite[see][]{O'Dwyer2010}, to the apparent brightening at the top of the arcade in response to the SAD collision (refer to figure~\ref{fig:sad_propagation_to_looptop_case3}), with the brightening at the ribbon location being seen through the top of the flare arcade. Further, since there is no significant change at the hot bump ($\log T=6.9$), the enhancement at the warm bump ($\log T=6.5$) cannot be explained by the cooling of the hot plasma (also evident from the animation associated with figure~\ref{fig:sad_case3_thermal_properties}, containing the EM[T] distribution for the aforementioned locations during 07:56:49--08:20:13 UT). Unlike SADs that occur earlier in the decay phase, the investigated SAD case does not carry 6 MK plasma, and hence cannot contribute directly to the EM enhancement. Thus, the EM enhancement at the warm bump is indeed due to the collision of SADs upon the flare arcade. Excluding with confidence the possibility of plasma rebound, we surmise that the collision is most likely magnetic in nature, considering the waves induced in the flare arcade and the associated footpoint brightening. However, we cannot exclude the possibility of mechanical heating at the top of the flare arcade, where the plasma could be compressed by the approaching SADs, resulting in the observed enhancement of the EM at $\sim$5 MK. On the other hand, we believe that it is unlikely that the plasma was heated to temperatures over 15 MK where AIA does not have sufficient sensitivity \cite[]{O'Dwyer2010}, because there are no appreciable changes of the EM component at $\sim\,$10 MK in response to the collision (Figure~\ref{fig:sad_case3_thermal_properties}). 
	
	Since the superposition of the top of the post-flare arcade with the upper footpoint can naturally boost the emission and the temperature at the upper ribbon, it may explain the consistently lower EM[T] values corresponding to the lower-ribbon location compared to that at the collision site (figure~\ref{fig:sad_case3_thermal_properties}f \& g). We also note in passing that in response to the SAD's interaction with the post-flare arcade, both the brightening in 1700~{\AA} and the EM enhancement at $\log T=6.7$ are much more pronounced at the upper (shadowed) ribbon than those at the lower (exposed) ribbon (figure~\ref{fig:sad_case3_thermal_properties}), which may indicate asymmetric heating and/or asymmetric loop geometry. Previous studies also report post-flare loops with one half of the loop being hotter and brighter than the other half \cite[see][and references therein]{Gou2016}. 
	
	Thus, our analysis indicates that the bright emission at the collision site generated in response to the SAD's arrival at and submergence below the top of the post-flare arcade exhibits the presence of hot plasma only up to 5 MK, whereas any substantial change of thermodynamical characteristics of plasma in the temperature larger than $\log T=6.7$ MK is not evident. It implies that the brightening at the collision site in projection can be contributed by both the heated plasma at the collision site of the SADs and the cooler plasma at the upper ribbon. The expansion of the latter may be responsible for the apparent ``rebound" motion discussed in \citet{Samanta2021}.
	
	\section{Discussion and Conclusion} \label{sec:disc_concl}
	Often observed during the post-maximum phase of the flare, supra-arcade downflows (SADs) are dark blob-like features that are believed to descend through the stretched cusp-shaped magnetic-field lines onto the post-flare loop arcade. This work presents the analysis of six episodes of SADs well observed during the decay phase of an M-class flare of April 11, 2013, which occurred in AR 11719 located close to the disk center. Primarily based on the analysis of limb events, previous investigations interpreted that SADs are either the cross-section \citep{McKenzie2000, Savage2010} or the wake \citep{Savage2012} of the retracting field lines. Alternately, SADs are attributed to the flow channels carved out by the reconnection generated outflow jets \citep{Cassak2013}. Here, we took advantage of the on-disk location of the flare to focus our investigation on probing the interaction of the SADs with the post-flare loop arcade and its low-atmosphere response, leading to our analysis and results differ substantially from those presented in \citet{Samanta2021} who investigated the SADs observed in the same flare.
	
	\subsection{Thermal characteristics of SADs}
	Thermal characteristics of the plasma contained in the SADs have been investigated since their earliest appearance. Our analysis revealed that SAD cases that occurred close to the flare-maximum contained hot plasma of temperature 5--7 MK (see figure~\ref{fig:sad_thermal_properties}). If the SADs were reconnection-generated sunward outflows \citep{Asai2004, Cassak2013}, the enclosed material could be attributed to the plasma heated during reconnection. However, the SAD cases which occurred in the late decay phase of the flare did not exhibit any discernible signature of hot plasma within. The presence of moderately hot plasma inside the SAD cases that occur close to the flare maximum despite its absence in the SAD cases which occur later in the gradual phase could mainly be attributed not only to the time duration passed since the flare peak, which is associated with diminishing reconnection rate, but also to the longer cooling times experienced by the SADs produced at higher and higher altitudes, thus explaining a variety of temperature and density of SADs obtained previously in the literature.
	
	It should be noted that our analysis indicates the presence of 5--7 MK plasma clearly only in 2 cases that occurred earlier, whereas, for the rest of the cases, the derived thermodynamical characteristics within and outside the SADs are similar to that obtained in \citet{Savage2012}. We further determined the level of depression of EM values in 7--12 MK at the SAD location relative to the surrounding, which indicates an increasing reduction (71\% to 43\%) in the high-temperature plasma content in the SADs that occur sequentially later in the decay phase. In comparison, \citet{Savage2012} estimated the depression of the EM values at the SAD location in 10--13 MK by a factor of 4 compared to that of the surrounding plasma. We note that the previously analyzed SAD cases often take place late in the flare decay phase, possibly limited by the observational contrast of SADs relative to the ambient bright supra-arcade region. For example, the SADs in \citet{Savage2012} occurred $\sim$90 minutes after the flare maximum. Recently, \citet{Brose2022} made the investigation of thermal characteristics of five SAD cases that occurred in the gradual phase of an M5.6 flare of 13 January 2015. In their SAD\#4 which occurred relatively early in the decay phase, they reported a localized enhancement of the EM-weighted temperature from 8.6 MK to 11 MK during the passing of the SAD. Therefore, more investigations are required for the SAD cases that occur in all stages of the flare decay phase to understand the comprehensive thermodynamical nature of the plasma contained in the SADs.
	
	\subsection{Waves \& oscillations induced by SADs}
	Our investigation of SADs' interaction with the structures in the supra-arcade region and post-flare loops revealed the generation of two distinct wave-like perturbations, namely the transverse oscillations exhibited by supra-arcade rays in the wake of the SAD passage (figure~\ref{fig:SAD_ray_transverse_osclln}), and downward propagating (speed $\sim$1000$\pm$500 km--s$^{-1}$) and expanding EUV intensity oscillations following the collision of the voids with the post-flare loop-arcade (figure~\ref{fig:sad_intensity_perturbation}). The periods of transverse oscillations of supra-arcade rays as well as those of intensity oscillations in the post-flare loop arcade are both found to be ranging between 120--150s. Since the propagating nature of transverse oscillations in the supra-arcade rays could not be confirmed, the oscillations are temporarily interpreted to be standing kink-waves \citep{Aschwanden2002}. Alternatively, \citet{Cooper2003} suggested that line-of-sight column depth variation of the coronal loops in the kink mode can result in intensity oscillations along the loop. On the other hand, the dynamical characteristics (e.g. period, speed) of intensity oscillations in the post-flare loop-arcade are similar to those obtained in the investigation of \citet{Verwichte2005}, who interpreted these sunward propagating waves (phase speeds up to 700 km--s$^{-1}$) as fast magneto-acoustic kink waves.
	
	\subsection{Loop foot-point brightenings associated with SADs}
	The interaction of the SAD with the apex of cusp-shaped loops in the supra-arcade region is temporally associated with the formation of kernel-shaped brightening at the foot-point region of the loops, as seen by combining 131{~\AA}, 1700{~\AA}, as well as H$\alpha$ images (case\#2 in Figure~\ref{fig:sad_propagation_to_looptop_case2} and case\#3 in Figure~\ref{fig:sad_propagation_to_looptop_case3}). By envisaging a cusp-shaped loop structure involved in the interaction with the SAD, the perturbations are estimated to propagate with a speed up to $\sim$600 km--s$^{-1}$ (Figure~\ref{fig:sad_propagation_to_looptop_case2} \& Figure~\ref{fig:sad_propagation_to_looptop_case3}). Observationally, such interactions appear to be similar to the cases in which SADs flow into the vertical current sheet from the edge-on perspective or into supra-arcade rays from the side-on perspective, resulting in the splitting of the latter structures as revealed in \citet{Innes2014}, and more recently in \citet{Liu2021}. Further, the rapid propagation of the energy, and the lack of evidence of localized strong heating induced by SADs at the `collision site' appear to be consistent with the picture proposed by \citet{Liu2013}, in which the dark voids (SADs) within the vertical current sheet (VCS) are mini flux-ropes formed during the process of reconnection. Subsequently, the highly stretched, post-reconnection flux tube smooths out its twist concentration, i.e., a plasmoid, as it retracts toward the post-flare arcade to become potential-like through Alfv\'en waves propagating towards the footpoints \citep{Fletcher2008}. Since in this picture the free energy (magnetic twist) carried by SADs is primarily transported to the surface by Alfv\'en waves and causes footpoint brightening, it is not expected to observe significant heating surrounding the SADs as found in our analysis.
	
	The collision of the SADs with the post flare loops resulted in intense EUV brightening over a relatively extended region at the collision site in projection, also reported previously by \citet{Samanta2021} who interpreted the apparent upflow of the bright material as rebounding motions. Our investigation of the ribbon response to the SAD's descent using 1700{~\AA} images (and in H$\alpha$ wavelengths) revealed co-temporal, co-spatial enhancements at the upper ribbon with a morphology similar to that observed at the collision site in the EUV images. Moreover, the speed of the ribbon expansion is found to be comparable to the apparent upflow speed of the EUV brightening observed after the collision. We further investigate the thermodynamical characteristics of the enhanced emission to discern the contribution of plasma at the collision-site from that at the ribbon. Since the upper ribbon is shadowed by the post-flare loop-arcade, we utilized enhancements observed in the lower (exposed) ribbon in 1700{~\AA} for this purpose. Emission measure distribution of the brightening generated at the collision site during and after the SAD's submergence below the post-flare loop top clearly indicates the enhancement of hot plasma only up to 5 MK, particularly in the temperature range $\log T\text{[K]}=[5.5,\ 5.9]$. This enhancement suggests that the apparent brightening at the top of the arcade in response to the SAD's arrival has a significant contribution from the enhanced ribbon emission in the transition region (Fe VIII at $\log T=5.6$; see \citet{O'Dwyer2010}). Thus, this analysis revealed that the brightening at the collision site is contributed not only by the heated plasma at the collision site of the SADs but also due to the footpoint emission shadowed by the flare arcade superposed along the line of sight. \citet{Samanta2021} found an increment in temperatures of the post-flare loop arcade to $\sim$20 MK in 10-15 minutes after the interaction of SADs. Further, recently \citet{Brose2022} investigated the nature of cooling of supra-arcade fans (SAFs) and coronal loops in a flare event with distinct episodes of SADs observed up to several hours after the impulsive phase. They found the observed cooling time of the SAFs and loops to be significantly longer than that estimated by the theoretical cooling model by \citet{Cargill1995}. However, the authors did not investigate how SADs affect the cooling of SAFs and loops in the gradual phase. Therefore, while the localized heating due to downflows has been amply reported in the previous investigations, the present work provides a definitive evidence of heating at the footpoint location associated with the SAD's descent.
	
\begin{figure}
		\centering
		\includegraphics[width=0.9\textwidth]{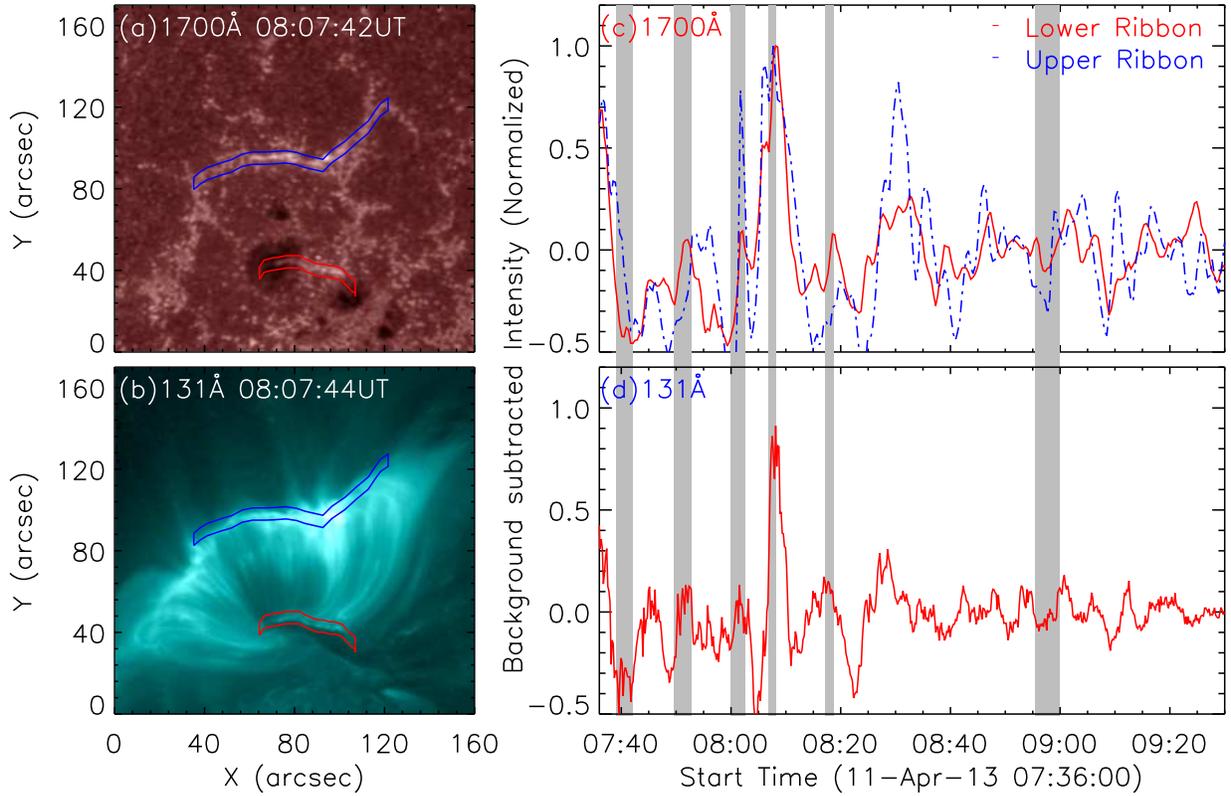}
		\caption{Emission at the ribbon location due to the SAD's interaction with the post-flare loop-arcade. Images in 1700{~\AA} (a) and 131{~\AA} (b) with the regions corresponding to the upper (lower) ribbon marked in blue (red), over which the intensity profiles are derived. Time profiles of emission in 1700{~\AA} (c) and 131{~\AA} (d), averaged over the marked regions, subtracted by a smoothed background of 20 minutes, and normalized to the peak value. Grey shaded regions correspond to the duration of different cases of SAD's propagation until submergence in the post-flare loops in the 131{~\AA} images.}\label{fig:ribbon_EUV_UV_brightening_profile}
	\end{figure}

	To further examine the association of ribbon brightening as seen in UV (1700{~\AA}) as well as in EUV (131{~\AA}) with the SAD propagation, we deduce the time evolution of averaged intensity over a 6" wide region spanning over the entire ribbon (figure~\ref{fig:ribbon_EUV_UV_brightening_profile}). We subtracted the deduced time-series with a background prepared by smoothing the time-series with a boxcar of 20-minute width. The normalized intensity profiles enable us to correlate the time periods of enhancements at the lower and upper ribbons with those of a SAD's descent until its submergence in the post-flare loops, marked by the grey shaded vertical regions. Except for cases \#1 and 6, we find that the foot-point brightenings are temporally linked with the SADs' interaction with the loops. This is in agreement with the detailed analysis of case\#2 and 3, presented in figure~\ref{fig:sad_propagation_to_looptop_case2} \& figure~\ref{fig:sad_propagation_to_looptop_case3}, respectively. For the cases where the association is not evident, we suspect that the interaction may not be strong enough. This analysis is similar to the investigation made in \citet{Asai2004}, where an excellent correlation is found between the downflows and non-thermal bursts in all the cases except one.  A periodicity of $\sim$10-minute is also found in the E/UV emission corresponding to the ribbon region, in agreement to that obtained by \citet{Samanta2021}, who, however, derived the intensities only in 131{~\AA} over the entire flare region. Therefore, our investigation agrees with \citet{Samanta2021} in the sense that the quasi-periodic nature of the EUV emission during the gradual phase of the flare might be regulated by descending SADs but we argue that this emission enlists a definitive contribution from the foot-point. The fractional contribution of the foot-point in the quasi-periodic EUV emission may have important implications for space weather yet remains to be investigated.
	
	Recently, \citet{Shen2022} investigated the nature of dark downflows in a three-dimensional magnetohydrodynamical model of solar flares and compared the model predictions with the nature of downflows as observed. Since our observations also contain the SADs above the arcade, we discuss the present results in context to the simulation of \citet{Shen2022}. Similar to observations, the finger-like voids in the simulation are developed in the interface region, i.e., the cusp or above-the-looptop region. However, the dynamic, bright finger-like structures above the loop top in the simulation seem to be different from the stable, extended spikes in observations \citep[e.g.,][]{Liu2021}. In their synthetic AIA 131~{\AA} image \citep[Fig. 2c in][]{Shen2022}, the termination shock front is still marginally visible as some diffuse, horizontal features; yet in observation, such features may become indiscernible against the stronger background on the disk than above the limb. On the other hand, the simulation does show that some dark voids penetrate deeply into the post-flare arcade and post-flare field lines are disturbed, and the disturbances do propagate along field lines to the surface (see their Supplementary Video 1), which are compared favorably to our observation.
	
	In conclusion, our analysis provides evidence that some SADs occurring close to the flare maximum may contain heated plasma, which may originate from their production sites, where magnetic reconnection is ongoing, and/or result from being compressed by the surrounding coronal plasma though which SADs descend. The combination of prolonged cooling and diminishing reconnection rate may explain the absence of such heated plasma in SADs produced in the late phase, which are associated with reconnections taking place in higher altitudes. The fact that intermittent SADs throughout the flare gradual phase may be associated with flare-ribbon emission enhancements corroborates a potentially important way of transporting flaring energy to the low atmosphere and signifies the role of SADs in quasi-periodic pulsation often observed during the flare gradual phase \citep{Hayes2019, Hayes2020, Zimovets2021}. Therefore, supra-arcade downflows are an excellent probe of the thermodynamical state of the plasma and energy release mechanism during the gradual phase of flares.
	
	\begin{acknowledgements}
	This work was supported by the National Natural Science Foundation of China (NSFC; Grant Nos 11950410498, 41774150, 11925302, 11903032, and 42188101) and by the National Astronomical Observatory of China (NAOC; Grant No KLSA202010). This work has received funding from the European Union's Horizon 2020 research and innovation program under the Maria Sklodowska-Curie grant agreement No. 847639. Authors thank the referee for help in improving the manuscript. Authors also acknowledge the open data policy of various space-based and ground-based observatories namely \textit{SDO}, \textit{GOES}, and GONG H$\alpha$ network. The use of the Helioviewer project, an open-source project for the visualization of solar and heliospheric data, is also acknowledged.
    \end{acknowledgements}
	
	\appendix
	
	\section{Thermal characteristics of the SAD cases 3, 4, and 6}\label{sec:appendix_1}
\begin{figure}
		\centering
		\includegraphics[width=0.73\textwidth]{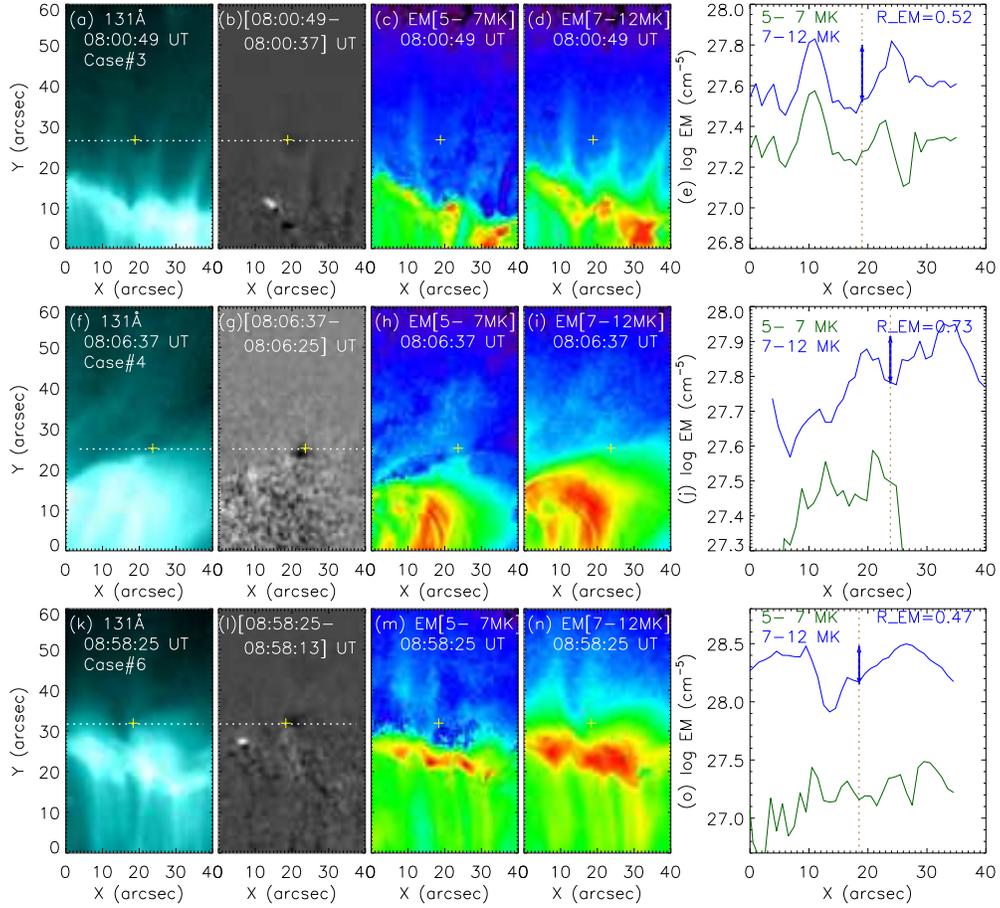}
		\caption{Thermal characteristics of the SAD cases 3, 4, and 6 in the top, middle and bottom panels, respectively. 131~{\AA} images (a) and the running difference image (b) has been utilized to identify SAD's location (shown by yellow '+' symbol). (c)--(d) EM[T] maps in 5-7 MK and 7-12 MK. (e) The evolution of EM[T] in 5--7 MK and 7--12 MK across the SAD (along the dotted horizontal line drawn in (a)\&(b)). Vertical double-sided arrow in (e) represents the level of depression (R\_EM), and the estimated value is also annotated. Similarly, middle and bottom panels show the thermal characteristics of the SAD case 4, and 6, respectively.}\label{fig:sad_thermal_properties_appendix}
	\end{figure}

	We have investigated the emission distribution (EM[T]) in all of the SAD cases that have been distinctively observed during the decay phase of the flare. While figure~\ref{fig:sad_thermal_properties} shows the analysis of thermal characteristics of the SAD cases 1, 2 and 5, the analysis for rest of the cases is presented in figure~\ref{fig:sad_thermal_properties_appendix}.
	
\end{document}